\documentclass[final,3p,times,twocolumn]{elsarticle}

\usepackage{amssymb}

\usepackage{psfig}
\usepackage{amsmath}
\usepackage{amssymb}
\usepackage{graphicx}

\newcommand{\lSect}[1]{{\label{sec:#1}}}
\newcommand{\lFig}[1]{{\label{fig:#1}}}

\def\gtaprx {\lower .1ex\hbox{\rlap{\raise .6ex\hbox{\hskip .3ex
    {\ifmmode{\scriptscriptstyle >}\else
        {$\scriptscriptstyle >$}\fi}}}
    \kern -.4ex{\ifmmode{\scriptscriptstyle \sim}\else
        {$\scriptscriptstyle\sim$}\fi}}}
\def\ltaprx {\lower .1ex\hbox{\rlap{\raise .6ex\hbox{\hskip .3ex
    {\ifmmode{\scriptscriptstyle <}\else
        {$\scriptscriptstyle <$}\fi}}}
    \kern -.4ex{\ifmmode{\scriptscriptstyle \sim}\else
        {$\scriptscriptstyle\sim$}\fi}}}
\newcommand{\note}[1]{\emph{\textcolor{red}{}}}

\newcommand{\Msun}{{\ensuremath{\mathrm{M}_{\odot}}}}

\newcommand{\FIGFF}[2]{{\ref{fig:#2}{#1}}}

\newcommand{\FIG}[2]{{Fig.~\FIGFF{#1}{#2}}}
\newcommand{\Fig}[1]{{\FIG{}{#1}}}

\newcommand{\Figure}[1]{{Figure~\FIGFF{}{#1}}}

\newcommand{\Sectff}[1]{{\ref{sec:#1}}}
\newcommand{\Sect}[1]{{\S~\Sectff{#1}}}

\newcommand{\Ep}[1]{{\ensuremath{10^{#1}}}}
\newcommand{\E}[1]{{\ensuremath{\powersep\Ep{#1}}}}

\newcommand{\powersep}{{\ensuremath{\times}}}
\newcommand{\cm}{{\ensuremath{\mathrm{cm}}}}
\newcommand{\erg}{{\ensuremath{\mathrm{erg}}}}

\newcommand{\CASTRO}{\texttt{CASTRO}}
\newcommand{\KEPLER}{\texttt{KEPLER}}

\journal{Computer Physics Communications}

\begin{document}

\begin{frontmatter}


 \title{Multi-Dimensional Simulations of Pair-Instability Supernovae}



\author [umn]{Ke-Jung Chen\corref{cor1}}
\author [umn]{Alexander Heger}
\author [lbl]{Ann S. Almgren }

\cortext[cor1]{Corresponding author; kchen@physics.umn.edu}
\address[umn]{School of Physics and Astronomy, University of
  Minnesota, Minneapolis, MN 55455,  United States}
\address[lbl]{Computational Research Division, Lawrence Berkeley National
  Lab, Berkeley, CA 94720,  United States}

\begin{abstract}
  We present preliminary results from multidimensional numerical
  studies of pair instability supernova (PSN), studying the fluid
  instabilities that occur in multiple spatial dimensions.  We use the
  new radiation-hydrodynamics code, \CASTRO, and introduce a new
  mapping procedure that defines the initial conditions for the
  multidimensional runs in such a way that conservation of physical
  quantities is guaranteed at any level of resolution.
\end{abstract}



\end{frontmatter}


\section{Introduction}
\label{intro}
The first stars that formed after the big bang may have a
characteristic mass 
scale 
of around hundred solar masses ($\Msun$) \citep{abel,bromm,larson}.
The stars with initial mass between $140\,\Msun$ and $260\,\Msun$ end
their lives in a very powerful explosion, called a pair-instability
supernova (PSN) \cite{barkat,bond,heger1}. During the evolution of those 
massive stars, after central carbon
burning, the core of the star reaches a sufficiently high temperature
that electron and positron pairs can be produced.  At this time,
radiation energy turns into rest mass for these pairs.  It softens the
adiabatic index $\gamma$ of the gas below the critical value of $4/3$
and triggers a rapid contraction that leads to explosive burning of
oxygen and silicon.  The energy released raises the pressure enough to
turn the contraction around into a energetic thermonuclear explosion
($\sim 3-100\E{51}\,\erg$).  These supernovae may play an important
role in the synthesis of heavy elements \citep{heger1,heger2} and
their energetic feedback to their surroundings can affect the
structure and evolution of the early universe \citep{carr,greif}.
Although several PSN candidates have been observed \citep{smith1,yam},
there are still many discrepancies between models and observations
\cite{kasen}.  The current theoretical models for PSN are all based on
one-dimensional calculations \cite{heger1,heger2}; until now, no
multidimensional simulations have scarce.  Here we study how
multi-dimensional fluid instabilities affect the mixing of elements
and possibly even the overall nucleosynthesis and energetics of PSN.

In this paper, we first introduce our numerical approach in
\Sect{castro}.  Then, in \Sect{mapping}, we introduce a new mapping
procedure that defines the initial conditions for the multidimensional
simulations in such a way that conservation of physical quantities is
guaranteed at any resolution.  We discuss the results of our
simulation in \Sect{results} and present our conclusions in
\Sect{conclusion}.

\section{Numerical Approach}
\lSect{castro} We start our simulations using one-dimensional models
obtained from the \KEPLER{} code \cite{weaver}, spherically symmetric
Lagrangian code that followed the evolution of a $150\,\Msun$ star up
to ten seconds before maximum compression.  Then we map the resulting
one-dimensional profiles into 2D and 3D to serve as the initial
conditions for \CASTRO{} \cite{ann1}.  We then evolve the simulation
for about one hundred seconds.  This is period during which
thermonuclear burning releases almost all of the energy of the
explosion.

\CASTRO{} \cite{ann1} is a new, massively parallel, multidimensional
Eulerian AMR radiation-hydrodynamics code for astrophysical
applications.  Time integration of the hydrodynamics equations is
based on a higher-order, unsplit Godunov scheme.  Block-structured
adaptive mesh refinement (AMR) and sub-cycling in time enable the use
of high spatial resolution where it is most needed.  We refer the
readers to \cite{ann1,ann2} for details.

\CASTRO{} allows the user to specify their equation of state, reaction
network, and method for calculating self-gravity.  For the simulations
presented here, we use the Helmholtz equation of state \cite{timmes}.
Our reaction network contains $19$ isotopes and $86$ reaction rates
\cite{weaver}.  This network is sufficient for us to model the burning
processes and their energy release to sufficient accuracy.  We use the
monopole approximation for self-gravity.

\section{Mapping of Initial Model in Multi-D}
\lSect{mapping} 

Since our initial model will be in a state close to hydrostatic
equilibrium, we need to be careful how we map the 1D spherically
symmetric data given on a non-uniform Lagrangian grid to a
multidimensional Eulerian grid.  Here we present a method that
numerically conserves quantities such as mass and energy that are
analytically conserved in the evolution equations.  Whereas this does
not guarantee that the initial data will be in perfect numerical
hydrostatic balance, it is at least a physically motivated constraint
and is sufficient for our simulations.  The algorithm as described
below is somewhat specific to our data set, but can be easily
generalized to other cases of mapping 1D data to higher dimensions.

Our original 1D data is given as cells of known size (radius and mass)
with known average values of intensive quantities (density, internal
energy).  Velocity is only given for the zone boundaries and hence
total momentum conservation is somewhat arbitrary\footnote{since the
initial model is spherically symmetric its total physical momentum
is always zero for every radius bin} as is kinetic energy
conservation.

The first step is to construct a continuous function (C$^0$) that
conserves the physically conserved quantities.
We found an ideal choice is to use ``\emph{volume coordinate}'', $V$,
the volume enclosed by a given radius from the center of the star.  We
then use densities (mass density, energy density) such that the
integral under the curve 
$$
X=\int_{V_1}^{V_2}\rho_X\,\mathrm{d}V
$$
is the total amount of the quantity $X$ with space density $\rho_X$
between the volume coordinates $V_1$ and $V_2$.

Here we use a piecewise linear function that preserves the original
monotonicity of the data (does not create artificial extrema) and is
bounded by the original extrema of the data (\Fig{detail}).  The
scheme causes some ``smearing'' (smoothing) of the data which,
however, is limited to less than one zone width of the original data.
We use a two-step process: First we construct a linear interpolation
across the interface between two zones which extends to the half-width
of the smaller of the two zones.  What is cut off from one zone by the
interpolation function ($a$ or $b$) is added to the interpolation in
the neighboring zone ($a'=a$ and $b'=b$).  This would usually result
in two ``kinks'' (change in slope) inside a zone (middle zone in
\Fig{detail}) with a flat piece that usually is a poor approximation
to the average gradient.  We now correct the interpolation within each
zone such that there is only one ``kink'' by finding a point in the
flat piece such that we have equal area enclosed by the triangles on
either side connecting from the new point to the value of the previous
interpolation function at the boundaries of the zone ($c=c'$).  

In principle one could now integrate this function over the volume of
the cell of the target grid to obtain the total amount of $X$ within
this cell.  For the sake of generality of the interpolation function,
we use an adaptive iterative sub-sampling method to obtain an
acceptably converged integral: We evaluate the interpolation function
at one point and multiply by the zone volume, then subdivide the zone
and sample the function for each of these sub-volumes, multiply by the
sub-volumes and sum up the result; this is repeated recursively for
each sub-zone until the results from the last two levels agree to
within the desired relative accuracy. Table 1 shows the results of mass 
deviation from original data after initial mapping by using 
linear interpolation and our algorithm; the number inside the () indicates 
the resolution of the mapped data. By comparing both results, our method has 
much higher accuracy and less depends on resolution and spacial dimensions. 

\begin{figure}
  \begin{center}
    \includegraphics[width=\columnwidth]{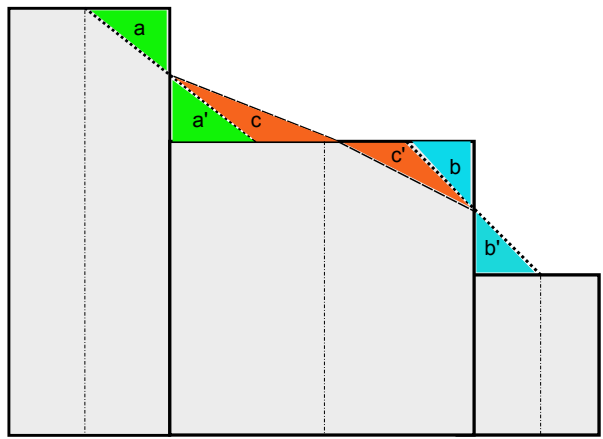}
  \end{center}
  \vspace{-\baselineskip}
  \caption{Schematic for constructing a conservative
    density profile.}
  \lFig{detail}
\end{figure}
This method is used to map internal energy density and the partial
densities of the chemical species.  The total density (or mass within
the zone) is then obtained from the sum of partial densities (masses);
pressure and temperature are obtained from the equation of state.
\begin{table}
\begin{center}
Table 1\\[2.0ex]
   \begin{tabular}{ | l | l | p{2.8cm} |} \hline
    Dimension &  Linear Interpolation  &  Our Algorithm  \\ \hline
1 D &    2.91$\%$(1024)  &   0.000254$\%$(1024) \\ \hline
2 D &    47.45$\%$($1024^{2}$)  &   0.005$\%$($1024^{2}$)\\ \hline
3 D &    127.05$\%$($128^{3}$)  &    1.85$\%$($2^{3}$)   \\ \hline
    \end{tabular}
\end{center}
\end{table}

\section{Results and Discussion}
\lSect{results} We performed our simulations on Calhoun at the
Minnesota Supercomputing Institute (MSI).  The 2D run used about
$4,000\,$CPUh; the 3D run used over $50,000\,$CPUh.  The results
presented here are from 2D runs of $150\,\Msun$ PSN in cylindrical
symmetry where we simulated only one hemisphere.

\Figure{PSN-e}{a} shows the specific nuclear energy generation rate 
in the inner $2\E{10}\,\cm$ domain at about $60$ secs after maximum 
compression. The color coding shows the specific nuclear
energy generation rate and in units of ergs/s/g on a logarithmic
scale.  We find Rayleigh-Taylor (RT) instabilities develop at the edge
of the oxygen-burning shell.  Later these instabilities will grow
further an affect such properties as the observable light curve of the
supernova.  \Figure{PSN-e}{b} is a close-up of the RT instability.

\section{Conclusions}
\lSect{conclusion} We have presented preliminary results from our
first multidimensional numerical study of the evolution of pair
instability supernova using the new Eulerian AMR
radiation-hydrodynamics code \CASTRO{}.  We simulated the formation of
Rayleigh-Taylor instabilities in the explosion of a $150\,\Msun$
pair-instability supernova.  We have introduced a new mapping method
that can be used to define the initial conditions for multidimensional
simulations from one-dimensional initial data in such a way that
conservation of physical quantities, monotonicity, and continuity are
guaranteed at any resolution.

\begin{figure}
  \begin{center}
    \includegraphics[width=\columnwidth]{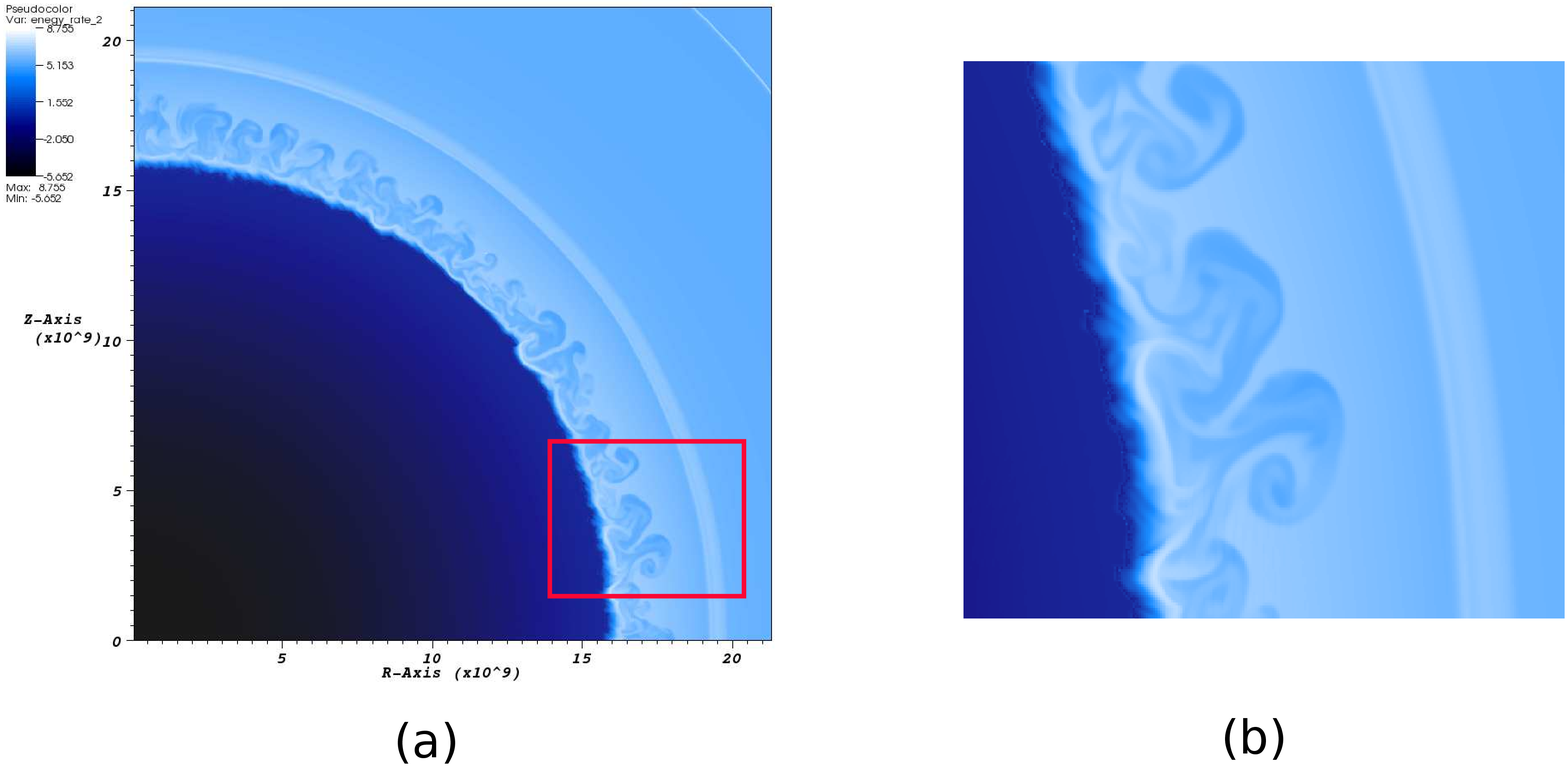}
  \end{center}
  \vspace{-\baselineskip}
  \caption{Energy generation rate\lFig{PSN-e}}
  \vspace{-\baselineskip}
\end{figure}

\section{Acknowledgments}
The authors would like to thank members of the Center for
Computational Sciences and Engineering (CCSE) at LBNL for their
invalueable support with using \CASTRO{}.  We would like to thank Candace Joggerst
and Haitao Ma for assistance with \CASTRO{} and Laurens Keek, David
Porter, Shuxia Zhang, Adam Burrows, and Stan Woosley for many useful
discussions.  The simulations were performed at Minnesota
Supercomputing Institute (MSI). This project has been supported by the
DOE SciDAC program under grants DOE-FC02-01ER41176,
DOE-FC02-06ER41438, and DE-FC02-09ER41618, and by by the US Department
of Energy under grant DE-FG02-87ER40328.

\section{References}
\bibliographystyle{cpc}
\bibliography{paper1_bibliography}






\end{document}